\documentclass[12pt]{article}

\usepackage{amssymb}
\usepackage{graphicx}
\usepackage{amsmath}
\usepackage{mathrsfs}
\usepackage{amsfonts}
\usepackage{cite}
\usepackage[hidelinks]{hyperref}

\begin{document}

%%%%%%%Author Note%%%%%%%%%%%%%%%%%%%%%%%%%%%%%%%%%%%%%%%%%%%%%%%%%%%%%%%%%%%%%%%%%%%%%%%%%%%%%%%%%%%

%Academia.edu Publishing LaTeX Template -- Please follow this optional template to format your article. You can also refer to article formatting instructions in the author guidelines
%For all sections, check the author guidelines to see if a specific section is required for your article type.

%\articletype{Research article}
%\proofstage{Template or Preprint}
%\jName{Academia Quantum}

%%%%%%%%%%%%%%%%%%%%%%%%%%%%%%%%%%%%%%%%%%%%%%%%%%%%%%%%%%%%%%%%%%%%%%%%%%%%%%%%%%%%%%%%%%%%%%%%%%%%%%

\title{Chaotic quantum transport through spatially symmetric microstructures 
in the symplectic ensemble}

%No Longer than 96 Characters (Inclusive of Spaces)

\author{Felipe Casta\~neda-Ram\'irez and Mois\'es Mart\'inez-Mares \\
Departamento de F\'isica, \\ Universidad Aut\'onoma Metropolitana-Iztapalapa, 
\\ Iztapalapa, Ciudad de M\'exico, Mexico}

\maketitle

%\affiliation{Departamento de F\'isica, Universidad Aut\'onoma 
%Metropolitana-Iztapalapa, Iztapalapa, Ciudad de M\'exico, Mexico.}

%\noindent $^{*}$Correspondence: Felipe Casta\~neda-Ram\'irez
%\email{$^{*}$felipe.castaneda.r@gmail.com}

\begin{abstract}
Quantum transport through left-right symmetric chaotic cavities in the presence 
of the symplectic symmetry, is studied through the statistical distribution of 
the dimensionless conductance. With this particular point symmetry, their 
associated scattering matrices are blocky diagonalized by a rotation by an 
angle of $\pi/4$. Although the formulation is established for an arbitrary 
number channels $N$, we present explicit calculations for $N=1$ and $N=2$, the 
last one showing the weak anti-localization phenomenon due to the symplectic 
symmetry.
\end{abstract}

%\keywords{chaotic quantum transport; scattering matrix; spatialy symmetric 
%cavities; symplectic ensemble.}

\section{Introduction}

Chaotic quantum devices have been of interest for more than thirty years, 
due to the fact that coherence gives rise to quantum interference that affect 
importantly their transport 
properties~\cite{Beenakker,MelloWRM,Alhassid,MelloBook,JPA38,Jalabert2016}. In 
that sense, the chaotic dynamics of the system makes the transport properties 
to fluctuate with respect to the Fermi energy~\cite{Keller}, an applied 
magnetic field~\cite{MarcusPRL69} or the shape of the system, or any other 
parameter that affect the dynamics of the carriers in the 
system~\cite{ChanPRL74}.

Of major importance in quantum transport is the role of symmetries, the 
presence or absence of time reversal symmetry for instance. A lot of work have 
been developed on this subject from both, theory and experiment, for quantum 
devices~\cite{Beenakker,MelloWRM,Alhassid,JPA38,MarcusPRL69,HuibersPRL81,
ChanPRL74,SantosKumar}, but also using auxiliary 
tools~\cite{MoisesPRE72,BaezPRE78,MoisesMelloPRE72}, like microwave  
cavities~\cite{SchanzePRE71,RafaelPIERS,YehPRE81,DietzPRE81} and 
graphs~\cite{SirkoPS2009,AngelPRB98,FelipePRE105}, and elastic 
systems~\cite{EnriqueSR}. In the absence of a magnetic field the time reversal 
invariance (TRI) for particles of integer spin, or half-integer spin and 
rotation symmetry, gives rise a phenomenon of weak 
localization~\cite{MarcusPRL69,Bergmann,BarangerPRL70}. For half-integer spin 
particles without rotation symmetry in the presence of TRI, the system exhibit 
the weak anti-localization phenomenon~\cite{Salawu}. None of these phenomena 
appear in the absence of TRI. In Dyson's scheme of symmetry 
classes~\cite{Dyson}, the first one corresponds to the orthogonal ensemble, the 
second one to the symplectic ensemble, and the last one to the unitary 
ensemble, respectively.

Apart from the symmetry classes, point symmetries have also been considered to 
observe interesting effects on the transport 
properties~\cite{BMPRB1996II,Kopp2008,WhitneyPRE,FelipeQR}. Of particular 
interest is the specular (left-right) symmetry 
with respect to a line that divide the two dimensional system into two parts, 
where the leads are symmetrically located, as well as its 
breaking~\cite{GMMBJPhysA1996,ZyczkowskiPRE56,MoisesPRE63,MoisesPRE71}. For 
this reflection symmetry, the statistical fluctuations of the dimensionless 
conductance were analyzed for the orthogonal and unitary symmetries; their 
generalizations to a spatial symmetry that corresponds to a rotation by an 
arbitrary angle in the Hilbert space, has been recently 
published~\cite{FelipeQR}. Such an analysis for the symplectic symmetry is 
still lack.

The purpose of the present paper is to obtain the statistical distribution of 
the dimensionless conductance of left-right symmetric cavities for the 
symplectic symmetry, when the leads support a few number of channels. In 
particular, the cases of one and two channels are explicitly shown. Also, 
we indicate how to extend these results to the generalized point symmetry 
through a rotation by an arbitrary angle in the Hilbert space.

We organize the paper as follows. In the next section we summarize the 
methodology used to describe quantum transport through left-right symmetric 
microstructures by means of the scattering matrix approach, we present useful 
representations of a scattering matrix, and the corresponding invariant 
measure. In section~\ref{sec:distribution} we apply this methodology to 
determine the distribution of the dimensionless conductance for the 
single channel case first, and then to the second channel case. Finally, we 
discuss our results in section~\ref{sec:discussion} .

\section{Specularly symmetric cavities in the symplectic ensemble}

\begin{figure}[!b]
\centering
\includegraphics[width=0.5\columnwidth]{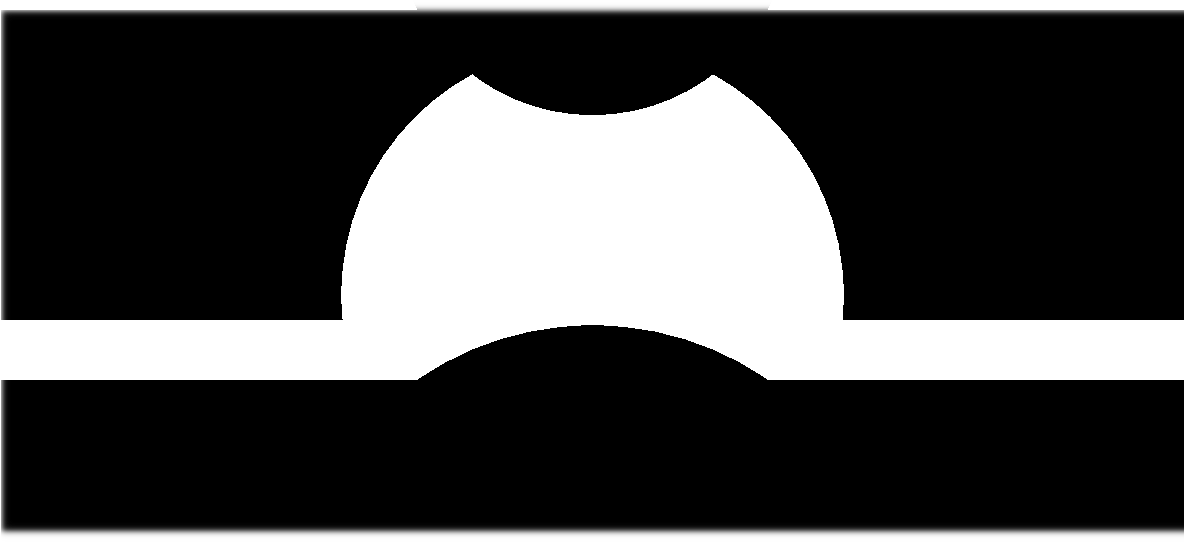}
\caption{A chaotic cavity with reflection symmetry connected to two waveguides 
symmetrically located.}
\label{fig:cavity}
\end{figure}  

Consider a chaotic cavity with reflection symmetry connected to two waveguides, 
as shown in Fig.~\ref{fig:cavity}. The two waveguides are symmetrically located 
and support $N$ channels each one, such that its associated scattering matrix 
is of dimension $4N$ and has the block symmetric structure
\begin{equation}
S = \left( 
\begin{array}{cc}
r & t \\ t & r
\end{array}
\right), 
\label{eq:SLR}
\end{equation}
where $r$ and $t$ are the reflection and transmission matrices, each one of 
dimension $2N$.

Scattering matrices with block symmetry are important to describe transport 
properties through scattering problems with specular symmetry in the absence 
of an applied magnetic field. Apart from the chaotic 
microstructures~\cite{GMMBJPhysA1996,BMPRB1996II,Kopp2008}, they have also 
appeared in wave transport systems, quantum or classical, in the context of 
disorderless lattices, like a linear chain of scatterers or a double Cayley 
tree~\cite{PRE2009,EPJST2017}. 

The symplectic symmetry impose the restriction to the scattering matrix to be 
self-dual; that is, $S^R=S$, where $S^R$ denotes the dual of $S$~\cite{Mehta}:
\begin{equation}
S^R=-Z_{4N}S^TZ_{4N} 
\label{eq:dual}
\end{equation}
where the superscript $T$ denotes transposition and $Z_{4N}$ is a block 
diagonal square matrix of dimension $4N$, with $2N$ matrices $z$ on the 
diagonal, where  
\begin{equation}
z = \left(
\begin{array}{cc}
0  & 1 \\
-1 & 0
\end{array}
\right).
\end{equation}

A scattering matrix with the structure of Eq.~(\ref{eq:SLR}) can be blocky
diagonalized by a $\pi/4$ rotation, such that it becomes parameterized in terms 
of two independent scattering matrices, $S_1$ and $S_2$, both of 
dimension $2N$. That is, 
\begin{equation}
S = R_{\frac{\pi}{4}}^T \left( 
\begin{array}{cc}
S_1 & 0_{2N} \\ 0_{2N} & S_2 
\end{array}
\right) R_{\frac{\pi}{4}}, 
\label{eq:Sdiag}
\end{equation}
where $0_{n}$ stand for the null matrix of dimension $n$ and
\begin{equation}
R_{\frac{\pi}{4}} = \left( 
\begin{array}{cc}
\mathbb{I}_{2N} & \mathbb{I}_{2N} \\ -\mathbb{I}_{2N} & \mathbb{I}_{2N}
\end{array}
\right) ,
\end{equation}
with $\mathbb{I}_{n}$ standing for the identity matrix of dimension $n$. A 
generalization of a diagonalization by a rotation by an arbitrary angle was 
recently proposed~\cite{FelipeQR}. The two matrices, $S_1$ and $S_2$, are most 
general scattering matrices that satisfies the condition (\ref{eq:dual}) for 
the symplectic ensemble: 
\begin{equation}
S_j^R=-Z_{2N}S_j^TZ_{2N}, \quad j=1,\, 2.
\label{eq:dual2}
\end{equation}

In terms of these two independent matrices, the reflection and 
transmission matrices are given by
\begin{equation}
r = \frac{1}{2}(S_1 + S_2) \quad\mbox{and}\quad 
t = \frac{1}{2}(S_1 - S_2).
\label{rts1s2}
\end{equation}
and the dimensionless conductance, obtained through 
$T=\mathrm{tr}(tt^{\dagger})/2$, according to the Landauer formula,  is given by
\begin{equation}
T = \frac{1}{8}\left[
4N - \mathrm{tr}\left( S_1S_2^{\dagger}+ S_1^{\dagger}S_2 \right)
\right].
\label{eq:Tdef}
\end{equation}

In the symplectic ensemble it is assumed that $S_j$ ($j=1,\,2$) belongs to the 
Circular Symplectic Ensemble, in which the scattering matrix has a uniform 
distribution. This uniform distribution is formally established through the 
invariant measure of the group $\mathrm{d}\mu(S_j)$. Then, the statistical 
distribution of $T$ can be obtained as
\begin{equation}
P(T) = \int \delta \left( 
T - \frac{1}{8}\left[
4N - \mathrm{tr}\left( S_1S_2^{\dagger}+ S_1^{\dagger}S_2 \right)
\right] \right)
\mathrm{d}\mu(S_1)\, \mathrm{d}\mu(S_2).
\end{equation}
Therefore, a parametric representation of the scattering matrices is 
needed.

\subsection{Representations of a scattering matrix}

Any self-dual unitary matrix $S_j$ ($j=1,\,2$) can be represented in terms of 
a most general $2N\times 2N$ unitary matrix $W_j$ as 
\begin{equation}
S_j = W_jW_j^R,
\label{eq:SUR}
\end{equation}
where $W_j^R$ is the dual matrix of $W_j$. Therefore, the invariant measure of 
$S_j$ is directly given by the invariant measure of $W_j$, 
\begin{equation}
\mathrm{d}\mu(S_j) = \mathrm{d}\mu(W_j).
\end{equation}

Another useful parameterization for $S_j$ is the polar 
representation~\cite{Beenakker}
\begin{equation}
S_j = \left( \begin{array}{cc} 
U_j & 0_{N} \\ 0_{N} & V_j 
\end{array} \right) 
\left( 
\begin{array}{cc} 
-\sqrt{1-\tau_j} & \sqrt{\tau_j} \\
\sqrt{\tau_j} & \sqrt{1-\tau_j} 
\end{array} \right) 
\left( 
\begin{array}{cc}
U_j^R & 0_N \\
0_N & V_j^R
\end{array} \right),
\label{eq:SS}
\end{equation}
where $\tau_j$ is a block-diagonal matrix whose diagonal elements are 
$\tau^{(j)}_1\mathbb{I}_2$, $\tau^{(j)}_2\mathbb{I}_2$,..., 
$\tau^{(j)}_N\mathbb{I}_2$, where $0\leq\tau^{(j)}_i\leq 1$ ($i=1,\ldots,N$) 
and $\mathbb{I}_2$ is the unit of dimension 2; $U_j$ and $V_j$ are arbitrary 
unitary matrices of dimension $N$. The invariant measure of $S_j$ is obtained 
from the invariant measures of $U_j$ and $V_j$, and the joint distribution of 
the $\tau$'s, given by~\cite{Beenakker}
\begin{equation}
p_j\left(\tau^{(j)}_1,\ldots,\tau^{(j)}_N \right) = C\prod_{i<j}^N 
\left|\tau^{(j)}_m-\tau^{(j)}_n\right|^4
\prod_{k=1}^N \tau_k
\label{eq:ptau4}
\end{equation}
That is,
\begin{equation}
\mathrm{d}\mu(S_j) = p_j\left( \tau^{(j)}_1,\ldots,\tau^{(j)}_N \right) 
\prod_{i=1}^n 
\mathrm{d}\tau^{(j)}_i \, 
\mathrm{d}\mu(U_j) \,
\mathrm{d}\mu(V_j).
\end{equation}

\subsubsection{Representation of SU(2) matrices}

Of particular interest is the representation of $2\times 2$ unimodular unitary 
matrix $U$. In the Hurwitz parameterization~\cite{Hurwitz}, $U$ can be 
written as~\cite{Karol}
 \begin{equation}
U = \mathrm{e}^{\mathrm{i}\phi}\, E, 
\label{eq:UCayley}
\end{equation}
where $E$ belong to the SU(2) group and has the structure
\begin{equation}
E = \left(
\begin{array}{cc}
a & b \\ 
-b^* & a^*
\end{array}
\right), 
\label{eq:SU2}
\end{equation}
with $a$ and $b$ being the Cayley-Klein parameters given by~\cite{Sakurai}
\begin{equation}
a = \mathrm{e}^{\mathrm{i}\zeta} \cos\varphi 
\quad\mbox{and}\quad
b = \mathrm{e}^{\mathrm{i}\chi}\sin\varphi 
\label{eq:Cayley-Klein}
\end{equation}
where $0\leq\varphi\leq\pi/2$, $0\leq\phi,\zeta,\chi\leq2\pi$. Its normalized
invariant measure is 
\begin{equation}
d\mu(U) = \frac{\mathrm{d}\phi}{2\pi}\, 
\mathrm{d}\mu(E) \quad\mbox{with}\quad
\mathrm{d}\mu(E) = 
\sin{(2\varphi)}\,
\mathrm{d}\varphi\,
\frac{\mathrm{d}\zeta}{2\pi}\, 
\frac{\mathrm{d}\chi}{2\pi}.
\label{eq:dmuE}
\end{equation}

Following Ref.~\cite{GMMBJPhysA1996}, we proceed to determine the 
statistical distribution of $T$ for few channels.

\section{Statistical distribution of the dimensionless conductance}
\label{sec:distribution}

\subsection{The single channel case}

When the waveguides support only one channel, the scattering matrix $S$ 
associated to the cavity is $4\times 4$. The scattering matrices $S_1$ and 
$S_2$ are $2\times 2$ and can be written as in Eq.~(\ref{eq:SUR}), where $W_j$ 
($j=1,\,2$) is a $2\times 2$ unitary matrix of the form of 
Eq.~(\ref{eq:UCayley}). At the end, $S_j$ becomes independent of $W_j$ because 
it reduces to 
\begin{equation}
S_j = \mathrm{e}^{2\mathrm{i}\phi_j}\, \mathbb{I}_2, \quad j=1,\,2.
\label{eq:Sj1ch}
\end{equation}
Also, the invariant measure of $S_j$ reduces to 
\begin{equation}
\mathrm{d}\mu'(S_j) = \frac{\mathrm{d}\phi}{2\pi} 
\int\mathrm{d}\mu(E_j) = \frac{\mathrm{d}\phi}{2\pi}
\label{eq:mu_1_Sj(1x1)}
\end{equation}
where $\phi_j$ is uniformly distributed in the interval $[0,2\pi]$. 

By replacing these matrices into Eq.~(\ref{eq:Tdef}) we may write $T$ as
\begin{equation}
T =\frac{1}{2}[1-\cos{(2\phi_1-2\phi_2)}],
\label{eq:Ttheta}
\end{equation}
This expression is the same as that found in Ref.~\cite{GMMBJPhysA1996} for the 
orthogonal and unitary symmetry classes, for which the statistical distribution 
of $T$ was obtained, with the result
\begin{equation}
P(T) = \frac{1}{\pi\sqrt{T(1-T)}}.
\label{eq:PTN1}
\end{equation}

\begin{figure}[!b]
\centering
\includegraphics[width=0.5\columnwidth]{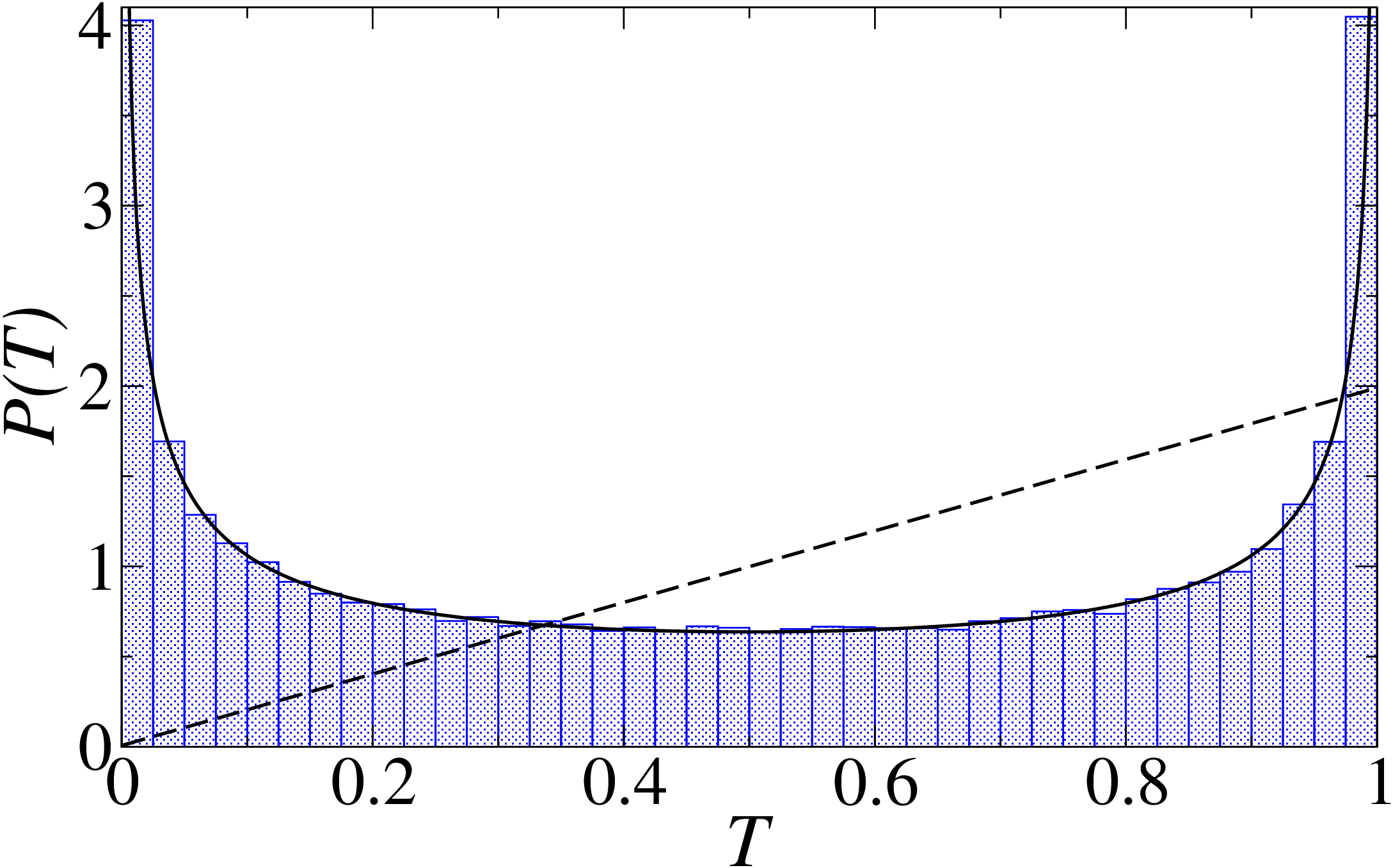}
\caption{Statistical distribution of $T$ for the single channel case. The 
histogram was obtained from a RMT simulation of $10^5$ realizations 
of scattering matrices of the form of Eq.~(\ref{eq:Sdiag}), where $S_1$ and 
$S_2$ are generated through their random phases uniformly distributed in the 
interval $[0,2\pi]$, see Eqs.~(\ref{eq:Sj1ch}) and (\ref{eq:mu_1_Sj(1x1)}). The 
continuous line is the theoretical result given by Eq.~(\ref{eq:PTN1}). The 
distribution of $T$ for asymmetric cavities is included for comparison 
(dashed line).}
\label{fig:PTN1}
\end{figure}

The histogram from the random matrix theory (RMT) simulation of $T$, obtained 
by generating numerically $10^5$ values of $\phi_j$ ($j=1,\,2$) with uniform 
distribution in the interval $[0,2\pi]$, is compared with the analytical result 
of Eq.~(\ref{eq:PTN1}) in Fig.~\ref{fig:PTN1}. 

\subsection{Two open channels}

In the case when the waveguides support two open channels, the scattering 
matrix becomes a $8\times 8$ matrix, and the scattering matrices $S_1$ and 
$S_2$ are $4\times 4$. For the latter matrices we may use the polar 
representation of Eq.~(\ref{eq:SS}) to write
\begin{equation}
S_j =
\left[
\begin{array}{cc}
-\sqrt{1-\tau_j}\, \mathrm{e}^{2\mathrm{i}\phi_j} \mathbb{I}_2 & 
\sqrt{\tau_j}\, \mathrm{e}^{\mathrm{i}(\phi_j+\psi_j)} D_j \\ 
\sqrt{\tau_j}\, \mathrm{e}^{\mathrm{i}(\phi_j+\psi_j)} D_j^R & 
\sqrt{1-\tau_j}\, \mathrm{e}^{2\mathrm{i}\psi_j} \mathbb{I}_2 
\end{array}
\right], \quad j=1,\,2,
\label{SS4x4}
\end{equation}
where $D_j$ ($j=1,\,2$) is a self-dual matrix, $D_j = E_j{E'_j}^R$, with $E_j$ 
and $E'_j$ defined as in Eq.~(\ref{eq:SU2}). 

Therefore, the expression for the dimensionless conductance is given by
\begin{eqnarray}
T(S_1, S_2) & = & 1 - 
\frac{1}{2}\sqrt{(1-\tau_1)(1-\tau_2)} 
\left\{\cos{[2(\phi_1-\phi_2)]} + \cos{[2(\psi_1-\psi_2)]}\right\}
\nonumber \\ & - & 
\sqrt{\tau_1\tau_2}\,
\alpha(E_1, E'_1, E_2, E'_2)\,
\cos{[(\phi_1-\phi_2)+(\psi_1-\psi_2)]},
\label{eq:Tp2ch}
\end{eqnarray}
where
\begin{eqnarray}
\alpha(E_1, E'_1, E_2, E'_2) & = & \alpha(\varphi_1, \varphi_1', \varphi_2, 
\varphi_2', \zeta_1, 
\zeta_1', \zeta_2, \zeta_2', \chi_1, \chi_1', \chi_2, \chi_2') 
\nonumber \\ & = & 
\cos(\zeta_1-\zeta_2-\zeta'_1+\zeta'_2) \cos{\varphi_1} 
\cos{\varphi_2} \cos{\varphi'_1} \cos{\varphi'_2} 
\nonumber \\ & + & 
\cos(\zeta_1-\chi_2-\zeta'_1+\chi'_2) \cos{\varphi_1}
\sin{\varphi_2} \cos{\varphi'_1} \sin{\varphi'_2}
\nonumber \\ & + & 
\cos{(\chi_1-\zeta_2-\chi'_1+\zeta'_2)} \sin{\varphi_1}
\cos{\varphi_2} \sin{\varphi'_1} \cos{\varphi'_2}
\nonumber \\ & + & 
\cos{(\chi_1-\chi_2-\chi'_1+\chi'_2)} \sin{\varphi_1}
\sin{\varphi_2} \sin{\varphi'_1} \sin{\varphi'_2}
\nonumber \\ & + & 
\cos{(\chi_1-\chi_2+\zeta'_1-\zeta'_2)} \sin{\varphi_1}
\sin{\varphi_2} \cos{\varphi'_1} \cos{\varphi'_2}
\nonumber \\ & - & 
\cos{(\chi_1-\zeta_2+\zeta'_1-\chi'_2)} \sin{\varphi_1}
\cos{\varphi_2} \cos{\varphi'_1} \sin{\varphi'_2}
\nonumber \\ & - & 
\cos{(\zeta_1-\chi_2+\chi'_1-\zeta'_2)} \cos{\varphi_1}
\sin{\varphi_2} \sin{\varphi'_1} \cos{\varphi'_2} 
\nonumber \\ & + & 
\cos{(\zeta_1-\zeta_2+\chi'_1-\chi'_2)} \cos{\varphi_1}
\cos{\varphi_2} \sin{\varphi'_1}\sin{\varphi'_2}.
\label{eq:g1}
\end{eqnarray}

In order to obtain the distribution of $T$, it is more convenient to determine 
the distribution of $T'=1-T$ first. The distribution of $T'$ is obtained as
\begin{equation}
P'(T') = \int 
\delta[T'-1+T(S_1,S_2)]\prod_{j=1}^2 p(\tau_j)\, 
\mathrm{d}\tau_j\, 
\frac{\mathrm{d}\phi_j}{2\pi}\, \frac{\mathrm{d}\psi_j}{2\pi}\,
\mathrm{d}\mu(E_j)\, \mathrm{d}\mu(E'_j)
\label{eq:pTp-1}
\end{equation}
where, see Eqs.~(\ref{eq:ptau4}) and (\ref{eq:dmuE}), $p(\tau_j)=2\tau_j$ and 
\begin{eqnarray}
\mathrm{d}\mu(E_j) & = & \frac{1}{2} \sin(2\varphi_j)\, 
\mathrm{d}\varphi_j\, \frac{\mathrm{d}\zeta_j}{2\pi}\, 
\frac{\mathrm{d}\chi_i}{2\pi}, \nonumber \\
\mathrm{d}\mu(E'_j) & = & \frac{1}{2} \sin(2\varphi'_j)\, 
\mathrm{d}\varphi'_i\,\frac{\mathrm{d}\zeta'_i}{2\pi}\, 
\frac{\mathrm{d}\chi'_i}{2\pi}, 
\label{eq:dmEEp}
\end{eqnarray}

\begin{figure}[!b]
\centering
\includegraphics[width=0.5\columnwidth]{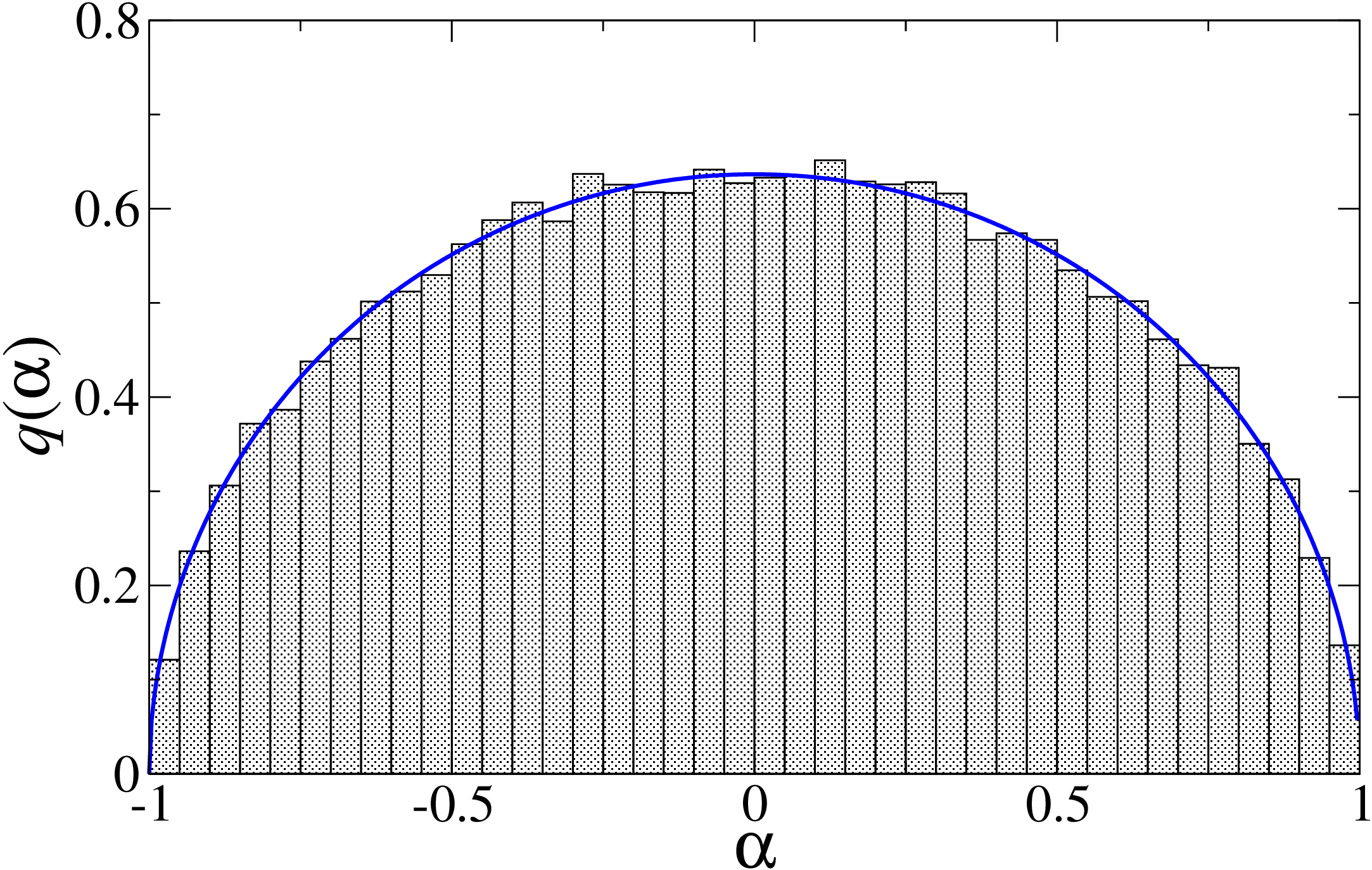}
\caption{Histogram of $10^5$ values of $\alpha(E_1,E'_1,E_2,E'_2)$ generated by 
RMT numerical simulation, compared with a theoretical guess given by 
Eq.~(\ref{eq:guess}) (continuous line).}
\label{fig:guess}
\end{figure}

Since the function $\alpha(E_1, E'_1, E_2, E'_2)$ that appears in the 
expression of $T(S_1,S_2)$ depends on the twelve parameters of the matrices 
$E_j$ and $E'_j$, for $j=1,\,2$, as can be seen in Eq.~(\ref{eq:g1}), it 
becomes advantageous to find its distribution to reduce the number of 
parameters to integrate in Eq.~(\ref{eq:pTp-1}). This is a complicated task to 
do analytically but we may explore the numerical simulation to propose the 
distribution of $\alpha$. The numerical simulation can be performed by 
generating random numbers according to the invariant measures of 
Eq.~(\ref{eq:dmEEp}). The result of this RMT numerical simulation is shown as 
histogram in Fig.~\ref{fig:guess}.

According to Fig.~\ref{fig:guess}, the distribution 
\begin{equation}
q(\alpha) = \frac{2}{\pi} \sqrt{1-\alpha^2}
\label{eq:guess}
\end{equation}
is a reasonable guess for the distribution of $\alpha\in[0,1]$. Therefore, 
\begin{equation}
P'(T') = \int 
\delta[T'-1+T(S_1,S_2)]\, q(\alpha)\, p(\tau_1)\, p(\tau_2)\,
\mathrm{d}\alpha\, 
\mathrm{d}\tau_1\, \mathrm{d}\tau_2\,
\frac{\mathrm{d}\phi_1}{2\pi}\, \frac{\mathrm{d}\phi_2}{2\pi}\,
\frac{\mathrm{d}\psi_1}{2\pi}\, \frac{\mathrm{d}\psi_2}{2\pi}
\label{eq:pTp-2}
\end{equation}

Except by the factor $\alpha$ in Eq.~(\ref{eq:Tp2ch}), the expression for 
$T(S_1,S_2)$ is the same as that for the orthogonal symmetry treated in 
Ref.~\cite{GMMBJPhysA1996}. Following that reference, $P'(T')$ reduces to 
\begin{eqnarray}
P'(T') & \propto & \int 
\delta\left[ T'- \frac{1}{2}\sqrt{(1-\tau_1)(1-\tau_2)} 
\left( \cos2\phi_1 + \cos2\psi_1 \right) - 
\sqrt{\tau_1\tau_2}\, \alpha\, \cos(\phi_1+\psi_1) \right]
\nonumber \\ & & \times
\sqrt{1-\alpha^2}\, \tau_1\, \tau_2\,
\mathrm{d}\alpha\, 
\mathrm{d}\tau_1\, \mathrm{d}\tau_2\, 
\mathrm{d}\phi_1\, \mathrm{d}\psi_1
\label{eq:pTp-3}
\end{eqnarray}
Now, a change of variables, $2\phi_1=\phi+\psi$ and $2\psi_1=\phi-\psi$, 
leads to 
\begin{eqnarray}
P'(T') & \propto & \int 
\delta\left[ T'- \frac{1}{2}\sqrt{(1-\tau_1)(1-\tau_2)} 
\cos\phi\cos\psi - \sqrt{\tau_1\tau_2}\, \alpha\, \cos\phi \right]
\nonumber \\ & & \times
\sqrt{1-\alpha^2}\, \tau_1\, \tau_2\,
\mathrm{d}\alpha\, 
\mathrm{d}\tau_1\, \mathrm{d}\tau_2\, 
\mathrm{d}\phi\, \mathrm{d}\psi.
\label{eq:pTp-4}
\end{eqnarray}
The integration with respect to $\alpha$ can be performed directly. We write 
the integral as
\begin{equation}
P'(T') \propto \int \left[ \int_{-1}^1 
\frac{\delta(\alpha - \alpha_0)}{\sqrt{\tau_1\tau_2}\cos\phi} 
\sqrt{1-\alpha^2}\, \mathrm{d}\alpha \right]
\tau_1\, \tau_2\,
\mathrm{d}\tau_1\, \mathrm{d}\tau_2\, 
\mathrm{d}\phi\, \mathrm{d}\psi,
\label{eq:pTp-5}
\end{equation}
where 
\begin{equation}
\alpha_0 = \frac{X}{\sqrt{\tau_1\tau_2}\cos\phi} , \quad\mbox{with}\quad
X = T'-\sqrt{(1-\tau_1)(1-\tau_2)}\cos\phi\cos\psi .
\end{equation}
Performing the integration with respect to $\alpha$ the remaining expression 
for $P'(T')$ in Eq.~(\ref{eq:pTp-5}) can be written as 
\begin{equation}
P'(T') \propto \int 
\frac{\sqrt{\tau_1\tau_2\cos^2\phi-X^2}}{\cos^2\phi} \, 
\mathrm{d}\tau_1\, \mathrm{d}\tau_2\, 
\mathrm{d}\phi\, \mathrm{d}\psi.
\label{eq:pTp-6}
\end{equation}
Lets define $\psi=\xi_1+\xi_2$, for $\xi_j\in[0,2\pi]$, and 
$\tau_j=\cos\theta_j$ for $\theta_j\in[0,\pi/2]$, for $j=1,\,2$. Then, we can 
write Eq.~(\ref{eq:pTp-6}) as
\begin{eqnarray}
P'(T') & \propto & \int_0^{\pi} \frac{\mathrm{d}\phi}{\cos^2\phi} 
\int_0^{\pi/2} \mathrm{d}\theta_1\, \cos\theta_1\sin\theta_1  
\int_0^{\pi/2} \mathrm{d}\theta_2 \cos\theta_2\sin\theta_2  
\nonumber\\ & \times & 
\int_0^{2\pi} \mathrm{d}\xi_1 
\int_0^{2\pi} \mathrm{d}\xi_2 \,
\sqrt{(Y_+\cos\phi-T')(Y_-\cos\phi+T') } \, ,
\label{eq:pTp-7}
\end{eqnarray}
where
\begin{equation}
Y_\pm=\cos\theta_1\cos\theta_2 \pm 
(\cos\xi_1\sin\theta_1\cos\xi_2\sin\theta_2 + 
\sin\xi_1\sin\theta_1\sin\xi_2\sin\theta_2).
\end{equation}

Defining the unit vectors
\begin{eqnarray}
\hat{\mathbf{u}}_1 & = & 
\cos\xi_1\sin\theta_1\, \hat{\mathbf{x}} + 
\sin\xi_1\sin\theta_1\, \hat{\mathbf{y}} + 
\cos\theta_1\, \hat{\mathbf{z}}, 
\nonumber \\ 
\hat{\mathbf{u}}_2 & = & 
\cos\xi_2\sin\theta_2\, \hat{\mathbf{x}} + 
\sin\xi_2\sin\theta_2\, \hat{\mathbf{y}} + 
\cos\theta_2\, \hat{\mathbf{z}}, 
\nonumber \\
\hat{\mathbf{v}} & = & - 
\cos\xi_1\sin\theta_1\, \hat{\mathbf{x}} - 
\sin\xi_1\sin\theta_1\, \hat{\mathbf{y}} + 
\cos\theta_1\, \hat{\mathbf{z}}.
\end{eqnarray}
where $\hat{\mathbf{x}}$, $\hat{\mathbf{y}}$, and $\hat{\mathbf{z}}$ are the 
canonical Cartesian unit vectors, we may have
\begin{equation}
\hat{\mathbf{u}}_1\cdot\hat{\mathbf{u}}_2 = Y_+, \quad 
\hat{\mathbf{v}}\cdot\hat{\mathbf{u}}_2 = Y_-, \quad 
\hat{\mathbf{u}}_1\cdot\hat{\mathbf{z}} = \cos\theta_1, 
\quad\mbox{and}\quad
\hat{\mathbf{u}}_2\cdot\hat{\mathbf{z}} = \cos\theta_2 \, ,
\end{equation}
such that Eq.~(\ref{eq:pTp-7}) is transformed to 
\begin{eqnarray}
P'(T') \propto \int_0^{\pi} \frac{\mathrm{d}\phi}{\cos^2\phi} 
\int \mathrm{d}\Omega_1\, \hat{\mathbf{u}}_1\cdot\hat{\mathbf{z}} 
\int \mathrm{d}\Omega_2\, \hat{\mathbf{u}}_2\cdot\hat{\mathbf{z}}
%\nonumber \\ & & \times 
\sqrt{(\hat{\mathbf{u}}_1\cdot\hat{\mathbf{u}}_2\cos\phi-T')
(\hat{\mathbf{v}}\cdot\hat{\mathbf{u}}_2\cos\phi+T')},
\end{eqnarray}
where $\mathrm{d}\Omega_j$ ($j=1,\,2$) is the solid angle element,
\begin{equation}
\mathrm{d}\Omega_j = \sin\theta_j\, 
\mathrm{d}\theta_j\, \mathrm{d}\xi_j 
\quad\mbox{for}\quad j = 1,\,2.
\end{equation}

Since the inner product is invariant under rotation we can make a rotation such 
that the unit vector $\hat{\mathbf{u}}_2$ remains along the $z$-direction. 
Therefore, the integration over $\Omega_2$ just is a constant and 
Eq.~(\ref{eq:pTp-7}) is simplified to 
\begin{equation}
P'(T') \propto \int_0^{\pi} \frac{\mathrm{d}\phi}{\cos^2\phi} 
\int_0^{\pi/2} \mathrm{d}\theta_1 \,
\cos\theta_1\sin\theta_1
\sqrt{\cos^2\theta_1\cos^2\phi-T'^2}.
\label{eq:pTp-8}
\end{equation}
The integration with respect to $\theta_1$ is direct if we make the change of 
the variable $\theta_1\to u=\cos^2\theta_1\cos^2\phi-T'^2$. The result can be 
written as
\begin{equation}
P'(T') \propto \int_0^{\pi/2} 
\frac{\left(\cos^2\phi-T'^2\right)^{3/2}}{\cos^2\phi} \, 
\mathrm{d}\phi,
\end{equation}
where we have considered that the integration from $\pi/2$ to $\pi$ is the 
same as that from 0 to $\pi/2$. 

Finally, we define the variable 
$\theta$ such that $\sin\phi=\sqrt{1-T'^2}\sin\theta$ and the integral is 
transformed to 
\begin{equation}
P'(T') \propto (1-T'^2)^2 \int_0^{\pi/2} 
\frac{\cos^4\theta}{\left[1-(1-T'^2)\sin^2\theta\right]^{5/2}}\, 
\mathrm{d}\theta. 
\end{equation}
The remaining integral that appear in this equation can be found in 
Ref.~\cite{Gradshteyn}, integral~(3.681.1). Therefore, the statistical 
distribution of $T$ is given by 
\begin{equation}
P(T) = \frac{1}{\pi}\, 
T^2(2-T)^2\,F\left[\frac{1}{2},\frac{5}{2};3;T(2-T)\right], 
\label{eq:PTN2b4}
\end{equation}
where $F(a,b;c;x)$ is the hypergeometric function and the normalization 
constant was obtained using the integral (7.512.4) of Ref.~\cite{Gradshteyn}.

\begin{figure}[!b]
\centering
\includegraphics[width=0.5\columnwidth]{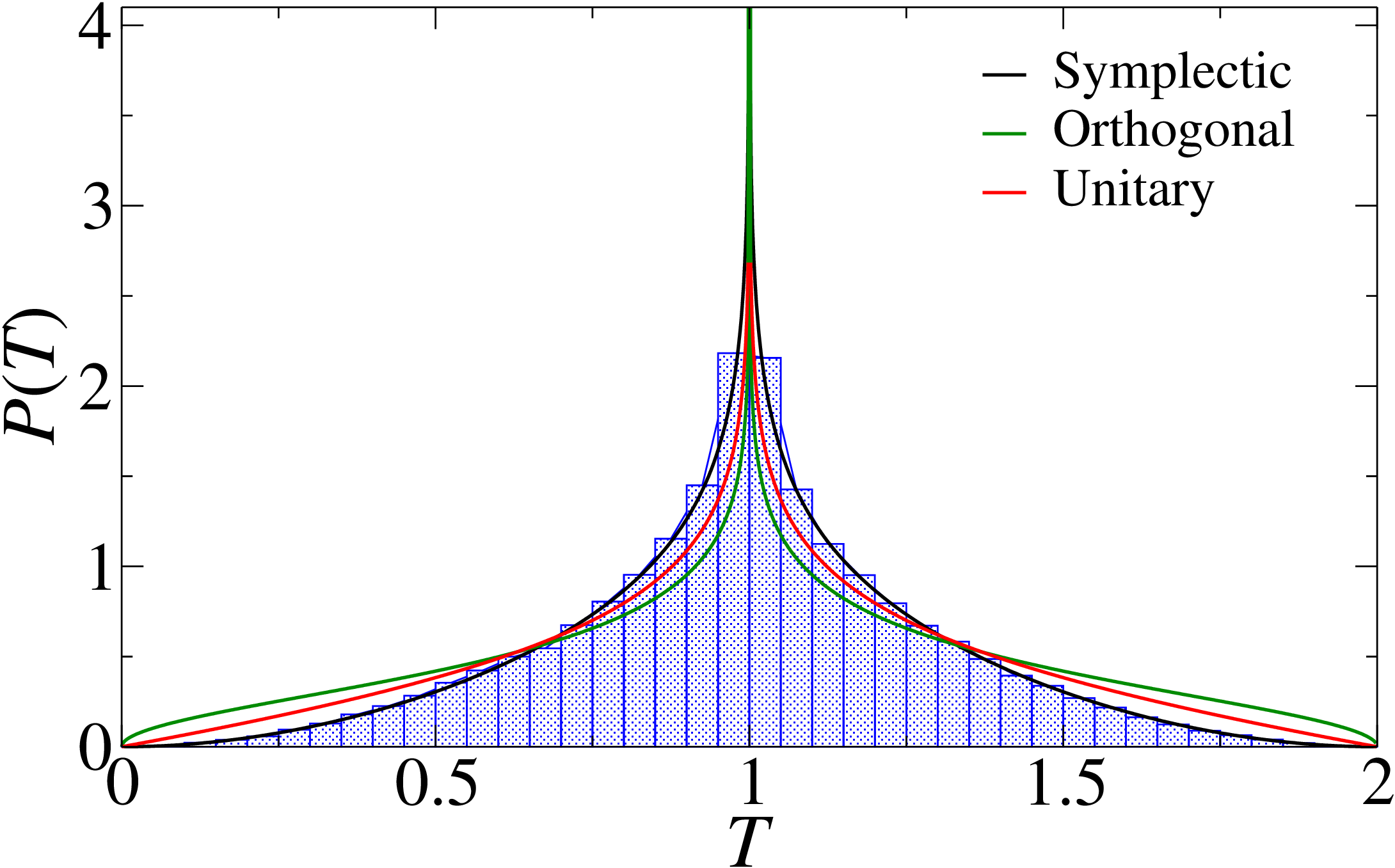}
\caption{Statistical distribution of $T$ for two channels on each lead. The 
histogram was obtained from a RMT simulation of $10^5$ realizations 
of scattering matrices of the form of Eq.~(\ref{eq:Sdiag}), with $S_1$ and 
$S_2$ generated using their polar representation, Eq.~(\ref{SS4x4}). The 
continuous black line is the theoretical result given by Eq.~(\ref{eq:PTN2b4}). 
For comparison, the known results for the orthogonal (green) and unitary (red) 
symmetry classes are included~\cite{GMMBJPhysA1996}.}
\label{fig:PTN2b4}
\end{figure}

In Fig.~\ref{fig:PTN2b4} we compare the result of Eq.~(\ref{eq:PTN2b4}) with 
the numerical simulation of $T$ obtained by generating $S_1$ and $S_2$ matrices 
in the form of the polar representation, Eq.~(\ref{SS4x4}), according to the 
respective invariant measure. 

\section{Discussion}
\label{sec:discussion}

A first important thing to note is the expression of the scattering matrix 
$S_j$ for the single channel case, Eq.~(\ref{eq:Sj1ch}). It is just a complex 
number of modulus 1;  except by a factor of $\pi/2$ the phase $\phi_j$ 
represents a phase shift due to a scattering potential or region (see 
Eq.~(2.297) of Ref.~\cite{MelloBook}). A similar expression is valid for the 
orthogonal and unitary symmetries. 

It is known that $1\times1$ scattering matrices distributed according to the 
invariant measure of the group, as in Eq.~(\ref{eq:mu_1_Sj(1x1)}), may 
describe chaotic cavities with a single mode port. It is clear that a 
$1\times1$ scattering matrix, does not reflect the presence of point symmetries 
in the cavity, coulding describe a symmetric or an asymmetric cavity. Moreover, 
being its invariant measure the same for the three symmetry classes, it 
indicates the independence on the symmetry class. This has a consequence for 
the two single mode ports case for the fluctuations of the dimensionless 
conductance. 

As can be seen in Fig.~\ref{fig:PTN1}, the effect of the left-right symmetry in 
the cavity is clearly observed through the statistical distribution of 
the dimensionless conductance in the one channel case, which is totally 
different for asymmetric cavities (dashed line in the figure). However, this 
distribution can not still discriminate the symmetry class because the matrices 
$S_1$ and $S_2$ do not depend on the symmetry class, as it was discussed in the 
previous paragraph.

For more than two channels, it is expected that the statistical fluctuations of 
the dimensionless conductance gives account of both, the symmetry class and the 
point symmetry. But this is the case of left-right symmetric cavities connected 
to two ports that support two channels each one. From one side, the analytical 
result for the dimensionless  conductance of symmetric cavities reflects the 
difference with respect to the asymmetric case~\cite{SantosKumar}. From other 
side, according to Fig.~\ref{fig:PTN2b4}, where the symplectic case is compared 
with the others two symmetry classes; a clear difference is observed. 
While the orthogonal symmetry shows the weak localization phenomenon, the 
symplectic symmetry presents a weak anti-localization phenomenon, both with 
respect to the unitary symmetry. 

Finally, for another point symmetry in which the scattering matrix of the 
system is blocky diagonalized by a rotation matrix of an arbitrary angle 
$\theta$, the statistical distribution of $T$ has the same structure as in 
Eq.~(\ref{eq:PTN2b4}), but replacing $T$ by $T/\sin^22\theta$~\cite{FelipeQR}.

\section*{Acknowledgments}

FC-R acknowledges financial support from Consejo Nacional de Humanidades 
Ciencia y Tecnolog\'ia (CONAHCyT), Mexico, under the scholarship with CVU 
number 885916.

\section*{Author Contributions}

Conceptualization, methodology, formal analysis, investigation and 
writing-original draft preparation were performed by both authors; 
writing-review and editing, supervision and project administration by MM-M. All 
authors have read and agreed to the published version of the manuscript.

\section*{Conflict of Interest}

The authors declare no conflict of interest.

%\section*{References}


\begin{thebibliography}{999}

% Reference 1
\bibitem{Beenakker}
C. W. J. Beenakker, 
Random-matrix theory of quantum transport, 
Rev. Mod. Phys. \textbf{69}, 731 (1997);
https://doi.org/10.1103/RevModPhys.69.731

%Reference 2
\bibitem{MelloWRM}
P.~A. Mello and H.~U. Baranger, 
Interference phenomena in electronic transport through chaotic cavities: an 
information-theoretic approach,
Waves Random Media \textbf{9}, 106 (1999);
10.1088/0959-7174/9/2/304

%Reference 3
\bibitem{Alhassid}
Y. Alhassid, 
The statistical theory of quantum dots, 
Rev. Mod. Phys. \textbf{72}, 895 (2000);
https://doi.org/10.1103/RevModPhys.72.895

%Reference 4
\bibitem{MelloBook}
P. A. Mello and N. Kumar, 
\emph{Quantum transport in mesoscopic systems: Complexity and statistical 
fluctuations} 
(Oxford University Press, New York, 2004).

%Reference 5
\bibitem{JPA38}
\emph{Special issue on trends in quantum chaotic scattering}, 
edited by Y. V. Fyodorov, T. Kottos, and H.-J. St\"ockmann, 
J. Phys. A: Math. Gen. \textbf{38} (49), 10433--10878 (2005).

%Reference 6
\bibitem{Jalabert2016}
R. A. Jalabert, 
Mesoscopic transport and quantum chaos,
Scholarpedia \textbf{11} (1): 30946 (2016).

%Reference 7
\bibitem{Keller}
M.~W. Keller, A. Mittal, J.~W. Sleight, R.~G. Wheeler, D.~E. Prober, R.~N. 
Sacks, and H. Shtrikmann, 
Energy-averaged weak localization in chaotic microcavities, 
Phys. Rev. B \textbf{53}, R1693 (1996).
https://doi.org/10.1103/PhysRevB.53.R1693

%Reference 8
\bibitem{MarcusPRL69} 
C.~M. Marcus, A.~J. Rimberg, R.~M. Westervelt, P.~F. Hopkins, and A.~C. 
Gossard, 
Conductance fluctuations and chaotic scattering in ballistic microstructures,
Phys. Rev. Lett. \textbf{69}, 506 (1992);
https://doi.org/10.1103/PhysRevLett.69.506

%Reference 9 
\bibitem{ChanPRL74} 
I.~H. Chan, R.~M. Clarke, C.~M. Marcus, K. Campman, and A.~C. Gossard, 
Ballistic conductance fluctuations in shape space,
Phys. Rev. Lett. \textbf{74}, 3876 (1995);
https://doi.org/10.1103/PhysRevLett.74.3876

%Reference 10
\bibitem{HuibersPRL81} 
A.~G. Huibers, S.~R. Patel, C.~M. Marcus, P.~W. Brouwer, C.~I. Duru\"oz, and 
J.~S. Harris, Jr., 
Distributions of the conductance and its parametric derivatives in quantum dots,
Phys. Rev. Lett. \textbf{81}, 1917 (1998);
https://doi.org/10.1103/PhysRevLett.81.1917

%Reference 11
\bibitem{SantosKumar}
S. Kumar and A. Pandey, 
Conductance distributions in chaotic mesoscopic cavities, 
J. Phys. A: Math. Theor. \textbf{43}, 285101 (2010);
stacks.iop.org/JPhysA/43/285101

%Reference 12
\bibitem{MoisesPRE72} 
M. Mart\'inez-Mares, 
Statistical fluctuations of the parametric derivative of the transmission and 
reflection coefficients in absorbing chaotic cavities,
Phys. Rev. E \textbf{72}, 036202 (2005);
10.1103/PhysRevE.72.036202

%Reference 13
\bibitem{BaezPRE78} 
G. B\'aez, M. Mart\'inez-Mares, and R.~A. M\'endez-S\'anchez,  
Absorption strength in absorbing chaotic cavities,
Phys. Rev. E \textbf{78}, 036208 (2008);
10.1103/PhysRevE.78.036208

%Reference 14
\bibitem{MoisesMelloPRE72} 
M. Mart\'inez-Mares and P.-A. Mello, 
Statistical wave scattering through classically chaotic cavities in the 
presence of surface absorption,
Phys. Rev. E \textbf{72}, 026224 (2005);
10.1103/PhysRevE.72.026224

%Reference 15
\bibitem{SchanzePRE71} 
H. Schanze, H.-J. St\"ckmann, M. Mart\'inez-Mares, and C. Lewenkopf, 
Universal transport properties of open microwave cavities with and without 
time-reversal symmetry,
Phys. Rev. E \textbf{71}, 016223 (2005);
10.1103/PhysRevE.71.016223

%Reference 16
\bibitem{RafaelPIERS} 
R.~A. M\'endez-S\'anchez, A.~M. Mart\'inez-Arg\"uello, G. B\'aez, and M. 
Mart\'inez-Mares, 
Scattering of waves: Imperfect coupling and absorption or amplification,
PIERS Proceedings (Moscow, Russia, 2012) p. 763.

%Reference 17
\bibitem{YehPRE81} 
J.-H. Yeh, J.~A. Hart, E. Bradshaw, T.~M. Antonsen, E. Ott, and S.~M. 
Anlage, 
Universal and nonuniversal properties of wave-chaotic scattering systems,
Phys. Rev. E \textbf{81}, 025201(R) (2010);
10.1103/PhysRevE.81.025201 

%Reference 18
\bibitem{DietzPRE81} 
B. Dietz, T. Friedrich, H.~L. Harney, M. Miski-Oglu, A. Richter, F. Sch\"afer, 
and H.~A. Weidenm\'"uller, 
Quantum chaotic scattering in microwave resonators,
Phys. Rev. E \textbf{81}, 036205 (2010);
10.1103/PhysRevE.81.036205

%Reference 19
\bibitem{SirkoPS2009} 
M. {\L}awniczak, S. Bauch, O. Hul1, and L. Sirko, 
Experimental investigation of properties of hexagon networks with and without 
time reversal symmetry,
Phys. Scr. \textbf{2009}, 014050 (2009);
10.1088/0031-8949/2009/135/014050

%Reference 20
\bibitem{AngelPRB98} 
A.~M. Mart\'inez-Arg\"uello, A. Rehemanjiang, M. Mart\'inez-Mares, J.~A. 
M\'endez-Berm\'udez, H.-J. St\"ockmann, and U. Kuhl, 
Transport studies in three-terminal microwave graphs with orthogonal, unitary,
and symplectic symmetry,
Phys. Rev. E \textbf{98}, 075311 (2018);
10.1103/PhysRevB.98.075311

%Reference 21
\bibitem{FelipePRE105} 
F. Casta\~neda-Ram\'irez, A.~M. Mart\'inez-Arg\"uello, T. Hofmann, A. 
Rehemanjiang, M. Mart\'inez-Mares, J.~A. M\'endez-Berm\'udez, U. Kuhl, and 
H.-J. St\"ockmann, 
Microwave graph analogs for the voltage drop in three-terminal
devices with orthogonal, unitary, and symplectic symmetry,
Phys. Rev. E \textbf{105}, 014202 (2022);
10.1103/PhysRevE.105.014202

%Reference 22
\bibitem{EnriqueSR} 
E. Flores-Olmedo, A.~M. Mart\'inez-Arg\"ello, M. Mart\'inez-Mares, G. B\'aez,
J.~A. Franco-Villafa\~ne, and R.~A. M\'endez-S\'anchez, 
Experimental evidence of coherent transport,
Sci. Rep. \textbf{6}, 25157 (2016);
doi: 10.1038/srep25157.

%Reference 23
\bibitem{Bergmann} 
G. Bergmann, 
Weak localization in thin films: a time-of-flight experiment with conduction 
electrons,
Phys. Rep. \textbf{107}, 1 (1984);
https://doi.org/10.1016/0370-1573(84)90103-0

%Reference 24
\bibitem{BarangerPRL70} 
H.~U. Baranger, R.~A. Jalabert, and A.~D. Stone, 
Weak localization and integrability in ballistic cavities,
Phys. Rev. Lett. \textbf{70}, 3876 (1993);
https://doi.org/10.1103/PhysRevLett.70.3876

%Reference 25
\bibitem{Salawu} 
Y.~A. Salawu1, J.~H. Yun, J.-S. Rhyee, M. Sasaki, and H.-J. Kim, 
Weak antilocalization, spin-orbit interaction, and phase coherence
length of a Dirac semimetal Bi$_{0.97}$Sb$_{0.03}$,
Sci. Rep. \textbf{12}, 2845 (2022);
https://doi.org/10.1038/s41598-022-06776-6

%Reference 26
\bibitem{Dyson}
F. J. Dyson, 
Statistical theory of the energy levels of complex systems. I, 
J. Math. Phys. \textbf{3}, 140 (1962);
 https://doi.org/10.1063/1.1703773 

%Reference 27
\bibitem{BMPRB1996II}
H.~U. Baranger and P.~A. Mello, 
Reflection symmetric ballistic microstructures: Quantum transport properties.
Phys. Rev. B \textbf{54}, R14297(R) (1996);
https://doi.org/10.1103/PhysRevB.54.R14297

%Reference 28
\bibitem{Kopp2008}
M. Kopp, H. Schomerus, and S. Rotter, 
Staggered repulsion of transmission eigenvalues in symmetric open mesoscopic 
systems,
Phys. Rev. B \textbf{78}, 075312 (2008);
https://doi.org/10.1103/PhysRevB.78.075312

%Reference 29
\bibitem{WhitneyPRE}
R.~S. Whitney, H. Schomerus, and M. Kopp, 
Semiclassical transport in nearly symmetric quantum dots. I. Symmetry breaking 
in the dot,
Phys. Rev. E \textbf{80}, 056209 (2009);
10.1103/PhysRevE.80.056209

%Reference 30
\bibitem{FelipeQR}
F. Casta\~neda-Ram\'irez and M. Mart\'inez Mares, 
Blocky diagonalized scattering matrices in chaotic scattering
with direct processes, 
Quantum Rep. \textbf{5}, 12 (2023);
https://doi.org/10.3390/quantum5010002

%Reference 31
\bibitem{GMMBJPhysA1996}
V.~A. Gopar, M. Mart\'inez, P.~A. Mello, H.~U. Baranger, 
The invariant measure for scattering matrices with block symmetries, 
J. Phys. A \textbf{29}, 881 (1996);
10.1088/0305-4470/29/4/014

%Reference 32
\bibitem{ZyczkowskiPRE56} 
K. $\dot{\mathrm{Z}}$yczkowsk, 
Scattering matrices with block symmetries,
Phys. Rev. E \textbf{56}, 2257 (1997);
10.1103/PhysRevE.56.2257

%Reference 33
\bibitem{MoisesPRE63} 
M. Mart\'inez and P.~A. Mello, 
Electronic transport through ballistic chaotic cavities: Reflection symmetry, 
direct processes, and symmetry breaking,
Phys. Rev. E \textbf{63}, 016205 (2000);
10.1103/PhysRevE.63.016205

%Reference 34
\bibitem{MoisesPRE71} 
M. Mart\'inez-Mares and E. Casta\~no, 
Effect of spatial reflection symmetry on the distribution of the parametric 
conductance derivative in ballistic chaotic cavities,
Phys. Rev. E \textbf{71}, 036201 (2005);
10.1103/PhysRevE.71.036201

%Reference 35
\bibitem{PRE2009}
M. Mart\'inez-Mares and A. Robledo, 
Equivalence between the mobility edge of electronic transport on disorderless 
networks and the onset of chaos via intermittency in deterministic maps,
Phys. Rev. E \textbf{80}, 045201(R) (2009);
10.1103/PhysRevE.80.045201

%Reference 36
\bibitem{EPJST2017}
M. Mart\'inez-Mares, V. Dom\'inguez-Rocha, and A. Robledo, 
Typical length scales in conducting disorderless networks,
Eur. Phys. J. Special Topics \textbf{226}, 417 (2017);
10.1140/epjst/e2016-60129-x

%Reference 37
\bibitem{Mehta}
M. L. Mehta,  {\em Random Matrices} (Elsevier Academic Press, New 
York, 1990) third ed. 

%Reference 38
\bibitem{Hurwitz}
A. Hurwitz,
\emph{Ueber die Erzeugung der Invarianten durch Integration} in 
\emph{Nachrichten von der Gesellschaft der Wissenschaften zu G\"ottingen, 
Mathematisch-Ph}, 
Nachr. K\"onigl. Ges. Wiss. G\"ott. Math.-phys. Kl. 
(Commissionsverlag del Dieterich'schen Universit\"atsbuchhandlung, G\"ottingen, 
1897) p. 71.

%Reference 39
\bibitem{Karol}
K. $\dot{\mathrm{Z}}$yczkowsk, 
\emph{Random Matrices of Circular Symplectic Ensemble} in \emph{Chaos -- The 
Interplay Between Stochastic and Deterministic Behaviour},  
Lecture Notes in Physics, vol 457, edited by P. Garbaczewski, M. Wolf, and A. 
Weron (Springer, Berlin, Heidelberg, 1995).

%Reference 40
\bibitem{Sakurai}
J.~J. Sakurai, 
\emph{Modern Quantum Mechanics}
(Addison-Wesley Publishing Company, Menlo Park, 1994) p. 170.

%Reference 41
\bibitem{Gradshteyn}
I. S. Gradshteyn and I. M. Ryzhik, 
\emph{Table of integrals, series, and products} 
(Academic Press, Orlando, 1980).

\end{thebibliography}
\end{document}